\documentclass{webofc}
\usepackage[varg]{txfonts}   
\newcommand{\Fig}[1]{Fig.~\ref{#1}}
\newcommand{\fig}[1]{\Fig{#1}}

\wocname{EPJ Web of Conferences}
\woctitle{Instabilities and Structures in Proto-Planetary Disks}
\begin{document}
\title{Elliptic and magneto-elliptic instabilities of disk vortices}
%
%

\author{Wladimir Lyra \inst{1,2,3}\fnsep\thanks{\email{wlyra@caltech.edu}}}

\institute{Jet Propulsion Laboratory, California Institute of Technology, 4800 Oak Grove Drive, Pasadena CA, 91109
\and Division of Geological \& Planetary Sciences, California Institute of Technology, 1200 E California Blvd MC 150-21, Pasadena, CA 91125
\and Sagan fellow
}

\abstract{%
 Vortices are the fundamental units of turbulent flow. Understanding
 their stability properties therefore provides fundamental insights on
 the nature of turbulence itself. In this contribution I briefly review the
 phenomenological aspects of the instability of elliptic streamlines, 
 in the hydro (elliptic instability) and hydromagnetic
 (magneto-elliptic instability) regimes. Vortex survival in disks is a
 balance between vortex destruction by these mechanisms, and
 vortex production by others, namely, the Rossby wave instability and
 the baroclinic instability.}
\maketitle
\section{Introduction}
\label{intro}

The first documented observation of non-laminar motion in fluids is present in 
the work of Leonardo da Vinci \cite{daVinci}. He called the phenomenon {\it  turbolenza}, 
after the Latin word for {\it swirl} (turbo). Sketching the flow, he wrote

\begin{quote}
  [...] the smallest eddies are almost numberless, and large things
  are rotated only by large eddies and not by small ones, and small 
  things are turned by small eddies and large.\\
\end{quote}

Though written around 1500, this passage reads surprisingly modern, containing the seeds of
concepts such as power spectrum and locality of the cascade. It 
contains also the insight that vortices are the fundamental unit 
of turbulent flow. The stability of vortices is thus a problem of 
paramount importance in fluid mechanics. Unveiling the mechanism 
that renders them unstable should provide vital insights into the
nature of turbulence itself. The instabilities of magnetized vortices 
should likewise provide a similar framework when it comes to MHD turbulence.

Phenomenologically, turbulence can be described as a series of
bifurcations, starting with a primary instability that converts shear
into vorticity, creating vortices. This is followed by another
bifurcation, a secondary instability, to break these vortices into lesser
vortical structures. These in turn shall experience a sequence of ``inertial                                                  
instabilities'', leading to a cascade. Though the Kelvin-Helmholtz instability
and the Rayleigh-Taylor instability are well established as examples
of primary instabilities, the highly successful theory of the turbulent
cascade put forth by Kolmogorov \cite{Kolmogorov} rested
on a heuristic picture of secondary instability, established
by early experiments \cite{Taylor}. It was not until the 80's
\cite{Pierrehumbert,Bayly} that the elliptic
instability was introduced as a mechanism for the secondary
instability. A fluid in rigid rotation supports a spectrum of
stable inertial waves, the simplest case being circularly
polarized transverse plane waves oscillating at twice the
frequency of the base flow \cite{Chandra}. Strain is introduced when the streamlines
pass from circular to elliptical, and some modes find resonance with
the strain field, leading to de-stabilization. 

The cascade reversal from inverse to direct when passing from 2D to 3D is a result of the different
properties of 2D and 3D vortices. Two-dimensional vortices do not 
decay, merging viscously and growing to the integral scale;
three-dimensional vortices also merge viscously, but before that
they generally fall prey to the elliptic instability, which does not
exist in two dimensions. Vortex survival depends on a balance between 
production and destruction. 

\section{Elliptic instability}

\begin{figure}
  \centering
  \sidecaption
  \resizebox{.5\textwidth}{!}{\includegraphics{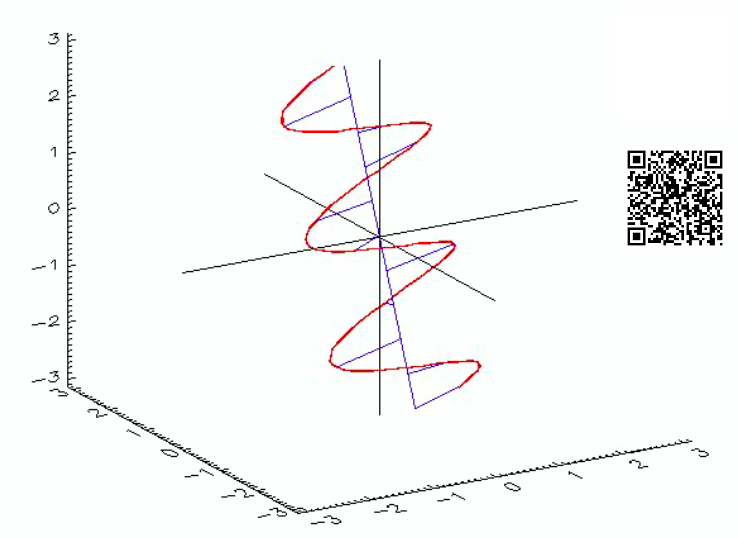}}
  \caption{Illustration of the circularly polarized tilted Kelvin
    wave, a mode supported by fluids in uniform rotation. The
    contour in red traces the wave. As the wave is transverse, fluid
    parcels execute motion along the blue ``arms'' of the figure,
    oscillating both in-plane and vertically. Scan the QR code for an 
    animation of the wave.}
\label{fig:helical-wave}       
\end{figure}

A column of fluid under rigid rotation supports stable oscillations in the form of
circularly polarized transverse plane waves (\fig{fig:helical-wave}). Restored by the Coriolis
force and propagating along a waveguide, these waves have also been
called {\it Kelvin waves}. As the waves are
transverse, if the direction of propagation coincides with the
rotation axis, the action of the wave is that fluid parcels will 
execute in-plane epicyclic oscillations. The propagation vector 
may also have an angle $\theta$ with the rotation axis, in which case the fluid 
motion is no longer in-plane, executing both epicyclic and vertical
motions, well known in galactic dynamics \cite{Binney-Tremaine}. 
Destabilization occurs when strain in introduced; 
steepening gradients and providing a source of free
energy. Instability occurs when a mode or pair of modes find resonance with the 
rotating strain field, which is to say when a multiple of the rotation 
frequency matches the frequency of the inertial waves.

\begin{figure}
  \centering
  \sidecaption
  \resizebox{.5\textwidth}{!}{\includegraphics{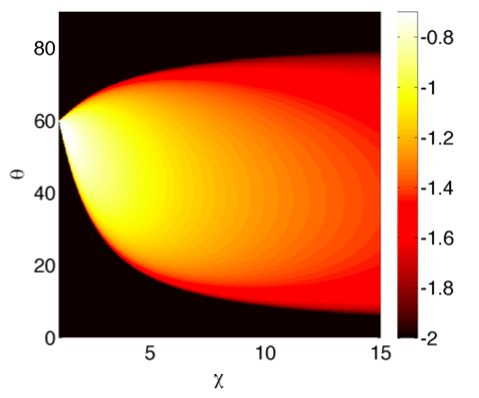}}
  \caption{Growth rates for perturbations in closed
    elliptic streamlines, in the simplest case of absent background rotation or 
    magnetic fields. In the $x$-axis
    $\chi$ is the aspect ratio of the base vortex, and $\theta$ the angle between the
    propagation vector of the inertial wave and the rotation axis of the
    vortex. There is no instability for circular streamlines
    ($\chi=1$): the instability is {\it elliptic}. No instability
    exists either for $\theta=0$, i.e., in-plane motion. The elliptic instability
    is inherently three-dimensional. Color-coded is the logarithm of
    the growth rate. Reproduced from \cite{Lesur09}.}
\label{fig:lesur}       
\end{figure}

The elliptic motion $U=\varOmega [-(1-\varepsilon)y, (1-\varepsilon)x]
$, where $0 \leq \varepsilon \leq 1$ is the ellipticity, is readily
decomposed $U=\varOmega (R+S)$ into rigid rotation $R=[-y,x]$ 
and the strain field $S=-\varepsilon[y,x]$. The growth rates, reproduced from 
\cite{Lesur09}, are shown in \fig{fig:lesur} in the $\chi$-$\theta$ plane. The aspect ratio $\chi$ of the streamlines 
is a measure of the strain. As the instability grows the vortex coherence is destroyed, energy
cascades forward and dissipates; the flow relaminarizes. 

As seen in \fig{fig:lesur}, the instability is {\em elliptical}:
there is no growth for $\chi$=1, i.e., circular streamlines in rigid rotation. The
instability is also inherently {\em three-dimensional}: there is no
growth for $\theta=0$, i.e., in-plane fluid oscillations. This later
behavior is at the heart of the difference between the cascade in 2D
and 3D. A stirring or primary instability, such as Rayleigh-Taylor or 
Kelvin-Helmholtz, generates the first eddies. The elliptic instability 
generates 3D turbulence out of this 2D motion, breaking the eddies. 
The growth rates are of the order of the turnover frequency, which
explains why vortices have lifetimes of this order. As elliptic
destruction occurs faster than viscous merging, the 3D cascade is
direct. In two dimensions, there is no elliptic instability, and the
eddies simply merge viscously. 

Adding a background rotation has significant effects for the
stability. The system now has two timescales: the turnover vortex 
time and that of the background flow. Not only that, the motions can
be either aligned (cyclonic vortex) or anti-aligned
(anti-cyclonic). Not surprisingly, the strongest instability occurs
for anticyclonic motion, as the two flows rotating in opposite
directions greatly enhances the effective shear on a fluid parcel. 
In this case, the in-plane, horizontal, motion gets de-stabilized. 
This instability is not of resonant nature, but centrifugal, appearing as exponential growth of
epicyclic disturbances. This behaviour invites a connection with the
Rayleigh (centrifugal) instability, and indeed the mechanism is
similar \cite{Lesur09}, suggesting that the Rayleigh instability is
a limit of the elliptic instability in the presence of rotation.

\section{Magneto-Elliptic Instability}

When magnetic fields are introduced in the problem, the addition of
Alfv\'en waves enriches the families of unstable modes. In the absence
of rotation, the magneto-elliptic instability is a parametric
instability as well \cite{Lebovitz-Zweibel}, have three unstable branches in
the $\chi$-$\theta$ plane. The first two are ``hydrodynamic'' and
``magnetic'', corresponding to resonant destabilization of Kelvin and
Alfv\'en waves, respectively. A third, ``mixed'' mode occurs through
instability of a pair of modes when one is hydrodynamic and the other
is magnetic, or a Kelvin-Alfv\'en instability. 

\begin{figure}
 \centering
 \sidecaption
  \resizebox{.5\textwidth}{!}{\includegraphics{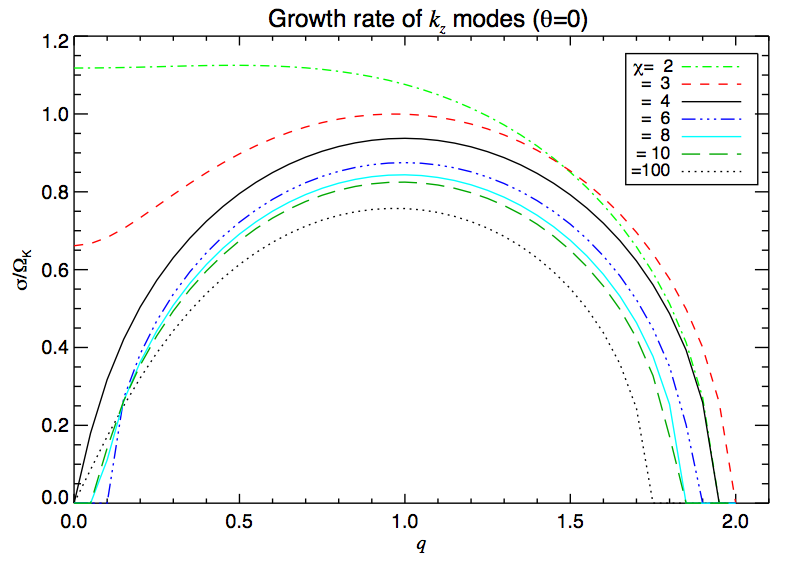}}
\caption{Growth rates of the horizontal ($\theta$=0) mode of the magneto-elliptic instability in the
  presence of rotation, for different aspect ratios of the base
  vortex. The quantity in the $x$-axis is $q=k/k_{\rm BH}$, where
  $k_{\rm BH}$ is the Balbus-Hawley wavenumber. As $\chi \gg 1$ we
  approach the pure shear limit ($\varepsilon = 1$) and the MRI growth
  rate is recovered: the MRI is a limiting case of the
  more general magneto-elliptic-rotational instability. Notice that
  for $q=0$ (no magnetic field), there is growth for $\chi=2$ and
  $\chi=3$, recovering the result of Rayleigh instability of
  horizontal modes of the hydro elliptic instability in
  the presence of rotation.  Reproduced from \cite{LyraKlahr11}.}
\label{fig:mri-ell}       
\end{figure}

When background rotation is introduced, the same as had occurred to the elliptic
instability ensues. When this background rotation runs opposite to the
rotation of the vortex (anti-cyclonic motion), a fluid parcel becomes subject to an intense
effective shear. Since the magnetic tension resists shear, leading to
instability 
\cite{Balbus-Hawley} a powerful unstable in-plane mode appears
\cite{Mizerski-Bajer}. 
This instability is of course the magneto-rotational instability, in
generalized form. Indeed, the dispersion relation of this horizontal
mode reduces to that of the MRI in the pure shear limit of
$\varepsilon=1$ (\fig{fig:rates}, \cite{LyraKlahr11,Mizerski-Lyra}). This
provides an interesting unification, explaining the
magneto-elliptic and magneto-rotational instabilities as
different manifestations of the same {\it magneto-elliptic-rotational
instability}.

\section{Formation and destruction of vortices in disks}

\begin{figure}
\begin{center}
  \resizebox{\textwidth}{!}{\includegraphics{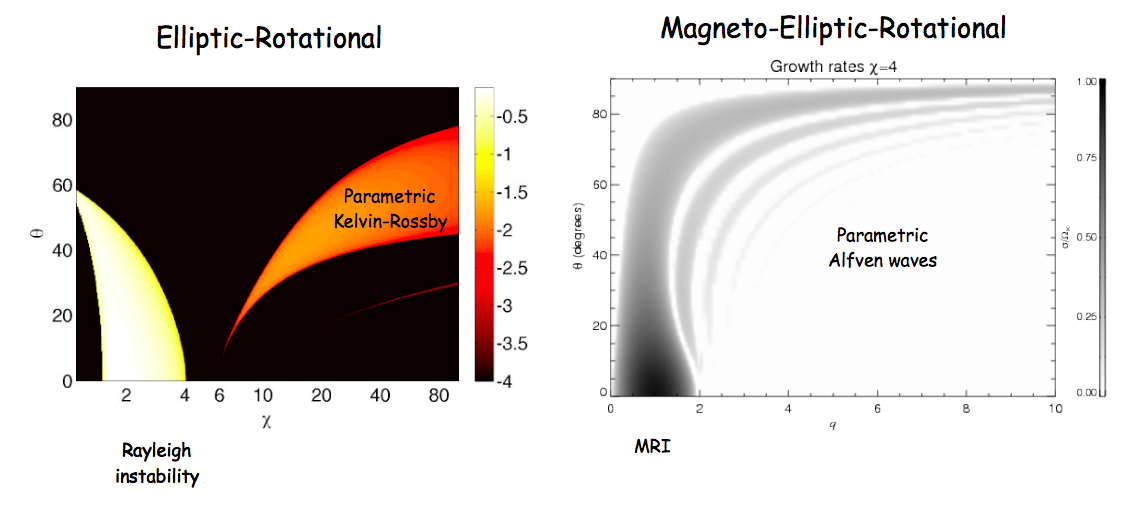}}
\end{center}
\caption{The four identified modes of vortex destruction. In the hydro
case (left panel, reproduced from \cite{Lesur09}), the Rayleigh
instability of horizontal modes, and the parametric resonant tuning of
3D inertial waves with the underlying strain field. In the magnetic
case (right panel, reproduced from \cite{LyraKlahr11}), the
magneto-rotational instability of horizontal modes, and resonant
tuning of Alfv\'en waves. These processes tend to oppose vortex
coherence, as growing perturbations extract kinetic energy from the vortical
motion. Vorticity is scattered, cascading forward, and eventually
being removed by viscosity. They thus behave as
vorticity sinks. A vortex survives if there is a mechanism to
counteract these losses, by injecting vorticity.}
\label{fig:rates}       
\end{figure}

\begin{figure}
\begin{center}
  \resizebox{\textwidth}{!}{\includegraphics{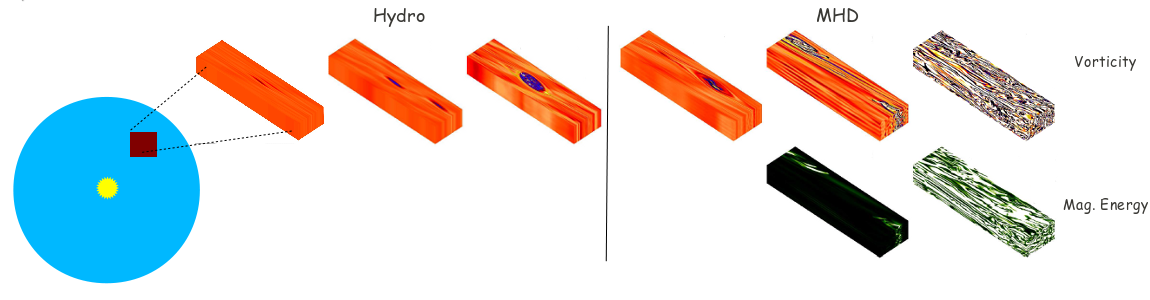}}
\end{center}
\caption{In non-magnetized flows, the vorticity injected by the
  baroclinic instability is able to counteract the vorticity lost the
  elliptic instability, at least the parametric, non-horizontal,
  version. In the magnetized case, however, the MEI growing inside the 
  vortex core is more than the vortex can withstand. The conclusion is
  that baroclinic vortices are restricted to the dead zones of
  accretion disks.} 
\label{fig:baro-mhd}       
\end{figure}

Four ways of destroying vortices in disks are identified in \fig{fig:rates}. In the
non-magnetic case (EI), these are the Rayleigh instability for the horizontal mode, and
the resonant destabilization of Kelvin waves. In the magnetic case
(MEI), these are the MRI for the horizontal modes, and resonant destabilization of 
Alfv\'en waves.

Counterbalancing these there are two mechanics to inject vorticity in
the flow. These are the Rossby wave instability (RWI, 
\cite{PapaloizouPringle,Lovelace99,Lyra08,Meheut,MKLin,Lyra-MacLow}), and the
Baroclinic Instability (BI,
\cite{Klahr04,Petersen07,Lesur10,LyraKlahr11}). 
The former is a linear instability powered
by a modification of the shear profile, that behaves as an external 
reservoir of vorticity. The latter is a nonlinear instability powered
by buoyancy and thermal diffusion, that establish a nonzero baroclinic
source term. 

As long as these vorticity sources (RWI and BI) are more powerful than
the vorticity sinks (EI and MEI), a vortex can survive. It has been
shown in the literature \cite{Lesur10,LyraKlahr11} 
that despite the elliptic instability, the baroclinic instability
keeps 3D vortices coherent. The result is core turbulence only, as the
vorticity lost by the EI is replenished by that injected by the BI.  

It has also been shown \cite{LyraKlahr11} that the same is not true
for the MEI, with a magnetized baroclinic vortex getting quickly
destroyed by the strong excitation of magneto-elliptic modes \fig{fig:baro-mhd}. Yet, 
magnetized vortices are seen in global, high-resolution
simulations \cite{Lyra-MacLow}. These were triggered by the
RWI, albeit artificially, due to an intense peak in magnetic pressure in the initial
condition, leading to a non-Keplerian shear profile in the active
zone. The cylindrical simulation was performed at very high resolution, $N_r,N_\phi,N_z$
= (768, 1536, 128) {\footnote{It also made use of high-end computing,
    being calculated in over 18\,000 processors at the NICS-Kraken
    cluster. We used the {\sc Pencil Code}, with which we achieve nearly linear
    scalability in NICS-Kraken, up to 70\,000 processors. }}
ranging $\pi$ in azimuth and 0.4 to 2.5 in
radius. This translates into 40 grid points per scale height, more
than enough to resolve the unstable magneto-elliptic modes. So, either
the said phenomenon is not a vortex but a zonal flow, or the RWI
provides vorticity faster than the MEI can destroy it, as is the case
with the BI and EI. If that is the case, we can group the processes in
order of strength as EI $<$ BI $<$ MEI $<$ RWI.

\end{document}